\documentclass{iau} 
\usepackage{graphicx}
\newcommand{\be}{\begin{eqnarray}}
\newcommand{\ee}{\end{eqnarray}}

\title[Observations meets theory in clustered star formation] 
{Observations meets theory in\\  clustered star formation}

\author[Susanne Pfalzner]   
{Susanne Pfalzner$^1$}
\affiliation{$^1$
J\"ulich Supercomputing Center, Forschungszentrum J\"ulich\\ 52428 J\"ulich, Germany \\ email: {\tt s.pfalzner@fz-juelich.de}
}

\pubyear{2019}
\volume{351}  
\jname{Star Clusters: From the Milky Way to the Early Universe}
\editors{A. Bragaglia, M.B. Davies, A. Sills \& E. Vesperini, eds.}
\begin{document}

\maketitle

\begin{abstract}
Stars form predominantly in groups which display a broad spectrum of masses, sizes, and other properties. Despite this diversity there exist an underlying structure that can constrain cluster formation theories. We show how combining observations with simulations allows to disclose this underlying structure. One example is the mass-radius relation for young embedded associations which follows $ M_c = CR_c^\gamma$ with $\gamma$ = 1.7 $\pm$ 0.2.0.2, which is directly related to the mass-radius relation of clumps.  Results based on GAIA 2D have demonstrated that young stellar groups (1—--5 Myr) expand and that this expansion process is largely over by an age of 10-20 Myr.  Such a behaviour is expected within the gas expulsion scenario. However, the effect of gas expulsion depends strongly on a the SFE, the gas expulsion time scale, etc. Here it is demonstrated how existing and upcoming data are able to constrain these parameters and correspondingly the underlying models.
\end{abstract}

\firstsection 
\section{Introduction}

Common knowledge from observations seems to be that most stars form in young clusters, which span a wide range of masses and sizes and mostly dissolve within approximately 10 Myr. Only the most massive clusters survive beyond that time. This information is usually applied to guide simulations of the development of young clusters. Here we have a closer look at these statements and their interpretation. What (additional) information do observations give us to guide theories of clustered star formation?

\section{The mass-radius relations}

We start with the statement that young clusters span a wide range of masses and sizes, which is definitely true. However, this is often interpreted that within reasonably bounds any combination of masses and sizes exist. For simulation that means either a single specific cluster is simulated or that a grid of initial conditions is applied in order to cover the supposed parameter space.

However, if we look at the actual observed parameters of young embedded clusters in the solar neighbourhood (see Fig. 1) a completely different picture emerges, namely, that these clusters do not cover the entire parameter space, but that a distinct correlation between mass and radius exists. This is roughly represented by

\be M_c = CR_c^\gamma \ee

with $\gamma$ = 1.7 $\pm$ 0.2 (\cite[Pfalzner et al. 2016]{Pfalzner_2015}). This means that clusters masses and sizes do span a wide range of masses and sizes, but not any combination is possible. It is essential that this is considered when setting initial conditions in cluster simulations that follow the dynamics of clusters in the embedded phase. Otherwise it is just an academic exercise that has little to do with the actual situation in real observed clusters.

Interestingly there exists a similar relation for the masses and sizes of clumps with approximately the same slope (Urquhart et al. 2014). This means the same applies for simulation of cluster formation. They have to start  out with mass-radius combinations that agree with these observations. 

\section{Two types of cluster}

Looking at clusters that are no longer embedded but have largely lost their gas, brings another surprise. There actually exist two types of clusters (\cite[Pfalzner 2009]{Pfalzner_2009}) and both develop along well defined tracks in the mass-radius-, and actually radius-age plane (see Fig. 1b). It has been suggested to distinguish them as clusters and associations (\cite[Gieles \& Portegies Zwart 2011]{Gieles_2011}), however, this leads to some confusion as many of the embedded "clusters" will dissolve as soon as they have lost their gas.  In the absence of a better naming convention, we will also refer to them as clusters and associations, but urge the community to clarify the nomenclature. The embedded clusters we have been looking at in Fig. 1 are actually the predecessors of the associations that largely dissolve within 5-10 Myr.

\begin{figure}[t]
\begin{center}
 \includegraphics[width=5.3in]{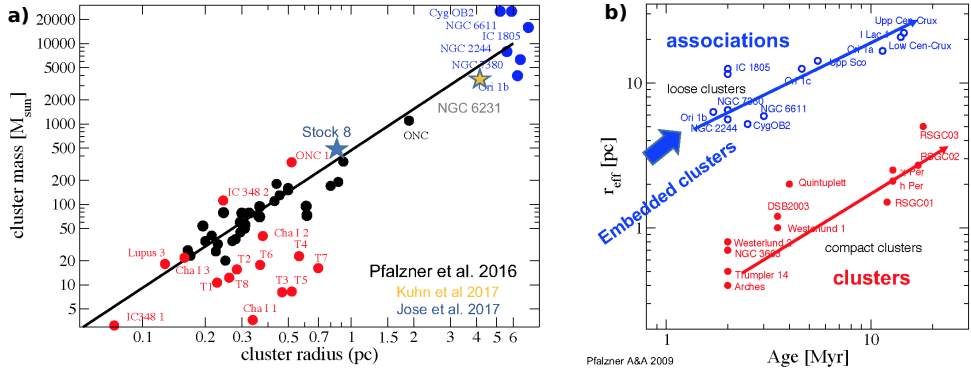} 
 \caption{Young cluster development a) Embedded associations, b) Expansion phase of clusters(red) and associations (blue).}
   \label{fig1}
\end{center}
\end{figure}

In each of the two classes the cluster size increases with age. However, the reason for this expansion differs in both cases significantly (\cite[Pfalzner \& Kaczmarek 2013]{Pfalzner_2013}). In associations the star formation is inefficient meaning that only a fraction ($\leq 30 \%$) of the gas and dust present in the clump is converted into stars.  The expansion is due to the supervirial state of the system when the gas has left the stellar group. 

Conversely, in clusters star formation is very efficient, with more than 60\% being converted into stars.  As such gas expulsion can not be the reason for their expansion. The reason why they expand nevertheless is that the stellar density in the cluster centre is so high that very close collision between stars are common. These interactions are so intense that stars are ejected at high speed from the centre of the cluster. As they come from deep within the potential well, the only way for the cluster to reach an equilibrium state again is expansion. Currently we have no observational information about the formation of the much more compact stellar groups that develop into long-lived clusters. It is one of the next big challenges to answer  the question how these compact long-lived clusters form.

\section{Cluster dynamics}

Therefore we concentrate in the following on the short-lived clusters that largely dissolve within 10 Myr. There has been a long debate about whether the more massive of these are formed in their currently observed state, or by the merging of sub-clusters, or expand. Recently, observations found that sub-clusters do \emph{not} merge, but instead 75\% of the investigated group of associations show strong features of \emph{expansion} (\cite[Kuhn et al. 2019]{Kuhn_2019}).

This means that theoretical models have to explain this expansion. Although there might be other models that can explain cluster expansion, the model that predicted this already a long time ago is that of gas expulsion. However, it is too simple an approach to speak of "the model of gas expulsion", because the outcome of gas expulsion simulations depends on many parameters. The main ones being:\\

\begin{itemize}
\item Star formation efficiency (for example, Tutukov 1978, Boily \& Kroupa 2003, Converse \& Stahler 2011)
\item Duration of gas expulsion phase (for example, \cite[Lada et al. 1984]{Lada_1984})
\item Constant vs density-dependent SFE  (\cite[Parmentier \& Pfalzner 2013]{Parmentier_2013})
\item Virial state before expulsion (for example, \cite[Allison \& Goodwin 2011]{Allison_2011}))
\item Spatial distribution before expulsion (clumping, central concentration) (for example, Fellhauer \& Kroupa 2005)\\
\end{itemize}

Now that GAIA DR2 data are availble one needs to deduce parameters from the observations that help to constrain better which of these models describe the observed clusters best. Basically one should provide developmental tracks for each of the different models, similar to those in provided by Fig. 2, to determine which model fits best the observational results.

\begin{figure}[t]
\begin{center}
 \includegraphics[width=3.3in]{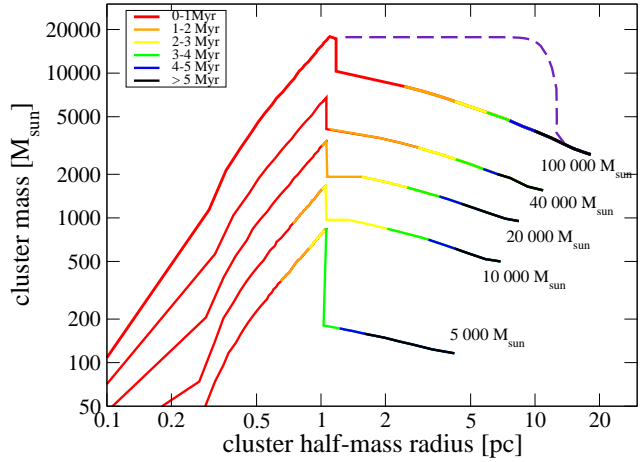} 
 \caption{Evelutionary tracks of young clusters according to  Pfalzner et al. (2015), assuming a star fromation efficiency that depends on the local gas density (Parmentier \& Pfalzner, 2014).}
   \label{fig1}
\end{center}
\end{figure}

\section{Observational constraints and evolutionary tracks}

 One can use the \cite[Kuhn et al. 2019]{Kuhn_2019} data to give better constraints on some of the parameters relevant for simulations (Pfalzner et al. in preparation). These data indicate that on average associations are fully embedded until about 1.4 Myr
and gas-free embedded at about 2.5 Myr. The length of the embedded phase seems to depend very little on 
associations mass. Conversely, the expulsion phases lasts longer for low-mass associations (2 Myr) than for 
high-mass associations(1 Myr).

However, these results should be regarded as preliminary, because the 25 stellar groups considered, do not only contain data of associations but also from clusters. As likely different mechanisms are responsible for their expansion, this might result in different expansion velocities. So future studies should avoid mixing of clusters and associations. 

The second problem, is that firm constraints require more associations/clusters to be studied to obtain reliable statistics.
So far the sample contains too few low-mass clusters at ages $>$ 3 Myr. The reason is that basically all known associations have been identified in the past as over-densities against the field population. The projected density of associations containing $\leq$ 1000 stars quickly falls below this detection limit as soon as associations start to expand (\cite[Pfalzner et al. 2015]{Pfalzner_2015}). This is why there are very few low-mass associations  older than 3 Myr identified. However, using the 3D information provided by Gaia, these associations should be identifiable. This would be really important to put better constraints on the development of low-mass clusters.

\section{Conclusion}
Observations demand that some of our assumptions have to be revised. Stars form in stellar groups which can be either associations or clusters. These two stellar groupings
both show distinct mass-size relations, which differ considerably. For associations the relation for the embedded phase should be used as determine the initial conditions for simulations.

Associations show clear signs of expansion, which currently is most easily explained as a result of gas expulsion, but alternatives explanations might exist. Most stars become unbound within 10 Myr, however, there remains a remnant which might be detectable in the Gaia DR2 data. In this context it is also necessary to look for new associations, to obtain better constraints on properties to test different models against.

Clusters survive because of their compactness, not their mass. Unlike for the associations we have no constraints on their embedded phase, so we    basically have no idea how they form. Obtaining data for this phase is one of the grand challenges of the future.

\end{document}